\documentclass[journal]{IEEEtran}
\usepackage{graphicx}
\usepackage{amsmath}
\usepackage{booktabs}
\usepackage{array}
\usepackage{textcomp}
\usepackage{url}
\usepackage{comment}

\begin{document}

\title{An Open Synthetic Test System for the Jordanian Transmission Grid}

\author{\IEEEauthorblockN{Muhy Eddin Za'ter\IEEEauthorrefmark{1} and
Bri-Mathias Hodge\IEEEauthorrefmark{1} \IEEEauthorrefmark{2}}

\IEEEauthorblockA{\IEEEauthorrefmark{1} School of Electrical and Computer Engineering,
University of Colorado Boulder, Boulder, CO 80305 USA}

\IEEEauthorblockA{\IEEEauthorrefmark{2} Renewable and Sustainable Energy Institute (RASEI),
University of Colorado Boulder, Boulder, CO 80309 USA}}

\maketitle

\begin{abstract}
Open synthetic test systems are essential for reproducible power system research, yet the available cases represent almost exclusively North American and European grids. No open transmission test system exists for any Middle Eastern country, whose grids raise different questions such as single corridor fuel-supply resilience and high-renewable operation within small synchronous systems, and established benchmarks cannot demonstrate. This paper presents a synthetic test system for the Jordanian transmission grid, assembled entirely from public sources, where the real topology is reconstructed from a published diagram and restored to its 2018 energized state, with plant-level generation and renewable fleets, per-site hourly profiles, and loads calibrated to the values published by Jordan's grid operators. The case is validated through structural statistics against real-grid criteria, power flow and $N\!-\!1$ screening, an energy-weighted loss decomposition, in addition to a full-year production-cost run compared against the published per-plant energy, and cross-solver verification. The model reproduces the annual energy mix at technology level (largest producers within 4\%, system total within 0.1\%), and the released dataset includes the bus-identity key, all scripts, and a post-2019 scenario variant. This test case is designed to benchmark system-level resource adequacy, time-series dispatch and unit commitment, renewable integration, fuel-supply resilience scenarios,  interconnection studies and steady-state studies.
\end{abstract}


\section{Introduction and Motivation}
\IEEEPARstart{A}{dvancement} in computational power systems research is highly benchmark-driven. Methods for unit commitment, economic dispatch, congestion management, contingency analysis, and renewable integration are developed, compared, and tested on a set of open test systems. Most notably the IEEE 14-, 30-, 57-, and 118-bus cases and their variants, such as the extended databases built upon the IEEE 118 bus~\cite{pena2017extended}. Beyond algorithm benchmarking, open test systems serve a second role as the basis for studies of a specific grid, including applications such as production cost modeling, resource adequacy assessment, renewable-integration planning, and transmission expansion. However, both of these roles are limited by the geographic coverage of the available test systems. The IEEE cases and their extensions mostly represent United States interconnections \cite{christie2000,pena2017extended,birchfield2016grid}, and many of the validated open power-flow models represent the European continental network \cite{hutcheon2013,horsch2018}. As a consequence, algorithmic development and system-specific studies are conducted on the topologies, fleet compositions, demand structures, and operating environments of these two regions, while grids elsewhere are effectively absent from reproducible research. Hence, a researcher studying a Middle-Eastern system must either adopt a foreign benchmark, where conclusions may not transfer to their regional grids, or work with confidential utility data that cannot be independently verified. To the authors' knowledge, no openly reproducible transmission test system exists for any grid in the Middle-Eastern region.

This gap is most significant where the unrepresented systems raise questions that the existing benchmarks are not designed to address, and the Jordanian system illustrates this gap. Jordan operates a relatively small, heavily fuel-import dependent power system in a conflict-afflicted region, and its recent history includes several external disturbances. For example, following the 2011 political instability in Egypt, which resulted in repeated attacks on the Sinai section of the Arab Gas Pipeline, in addition to a shortfall in Egypt's own domestic gas, there were interruptions to the natural gas imports that had supplied the bulk of Jordanian generation \cite{alshwawra2023electricity}. This forced the system towards emergency oil firing~\cite{schuetze2023} before liquefied natural gas imported through a terminal at Aqaba restored an 88\% gas share by 2018~\cite{nepco2018}. Additionally, the interconnection with Syria, which since 2001 has served as a synchronization and an exchange corridor for the northern Jordanian subsystem, was disrupted by regional conflict and has carried no power exchange since 2012~\cite{nepco2018}. 
The Syrian crisis also shaped the fleet in which the engine plants IPP3 (2014) and IPP4 (2015) were procured in this period because they can run on gas, heavy fuel oil, or diesel.
In parallel, the Jordanian grid is rapidly decarbonizing; the renewable capacity grew from approximately one gigawatt in 2018 to more than double by 2020, exceeding a 20\% energy share~\cite{batarseh2022covid}. 
%The demand environment in Jordan is dominated by occupancy-driven load (residential consumption alone accounts for 45\% of the total, with commercial use adding a further 15\%~\cite{nepco2018}), with only 21\% industrial load, and the remainder of the load is attributed to agriculture, water pumping (15\%) and street lighting~\cite{nepco2018}. 
% This load mixture creates a distinct set of operational characteristics, where a time-shifting morning minimum load under rising solar output, a widening peak-to-valley spread, and a load factor that declined from 65\% toward 60\% over 2019--2020~\cite{batarseh2022covid}. The representative weeks of Fig.~\ref{fig:prof} show these features in the synthesized 2018 load time series. 
Also, the demand is dominated by occupancy-driven residential and commercial load with a small industrial base, a composition detailed in Section II-A.

Each of the above mentioned characteristics corresponds to an active research area. For example, the interdependence of gas supply and electric operation is an established field~\cite{shahidehpour2005}, as is power-system resilience to external disruption~\cite{panteli2015} and the operation of small synchronous systems under high renewable penetration~\cite{osullivan2014}. Studies in these areas require a test system whose structure contains the characteristics under study (i.e fuel supply structure, interconnection strength and renewable generation location).
These factors are fixed by the choice of benchmark, and none of the available cases combines a single dominant import corridor, an effectively isolated synchronous system, and renewable capacity concentrated at the far end of a long network.
% Available benchmarks typically assume large, interconnected system with diversified fuel supplies, conditions that different from the Jordanian grid. Where Jordan relies on a single dominant fuel delivered through one import corridor, operates as an effectively isolated synchronous system, and concentrates its renewable capacity at the far end of a long transmission network. Therefore, studying issues like supply resilience, fuel-electric interdependence, and high-renewable operation within a small synchronous system requires a dedicated representation of the Jordanian grid.

These considerations motivated the presented model in this work, in which fuel supply, interconnections, and fleet availability are represented explicitly so that such disturbances can be formulated as input changes. Assembling such a model from public sources is feasible for Jordan. As NEPCO publishes detailed annual system accounts~\cite{nepco2018}, a national master-plan study documents the network and the generation fleet~\cite{jica}, a published PMU placement study provides a numbered single-line diagram of the transmission network~\cite{al2023global}, and an analysis of the demand is available in~\cite{batarseh2022covid}.

The primary contribution of this work is the dataset itself. The methods of creating the dataset are motivated and verified in the works of of~\cite{pena2017extended, birchfield2016grid, birchfield2018techniques}. Like the NREL-118 database, the model does not claim to be the operational NEPCO network, it is a synthetic representation that includes a reconstructed topology restored to its 2018-energized state using dated public records. Alongside computed and standard electrical parameters, a plant-level renewable generation fleet is provided with hourly profiles derived per site from open satellite-irradiance and reanalysis databases (NSRDB, MERRA-2). Loads are calibrated to the published system totals (peak demand, annual energy, and load factor). The case is built as a 2018 snapshot, for which a complete public calibration set exists, and is validated using structural statistics compared against real-grid criteria as outlined in \cite{birchfield2016grid}. Operational validation is also conducted including running power flow and $N\!-\!1$ screening, an energy-weighted loss decomposition against the metered annual figure, and a full-year production-cost run in which the annual energy mix is compared to the published per-plant accounts. The released model comprises the network in open formats with cross-solver verification, the bus-substation mapping key, all data generation scripts, and a post-2019 topology variant. Therefore, every reported result is reproducible from public inputs and the parameters of the above paragraph are directly editable. The model and the scripts are available here link\footnote{https://github.com/HodgeLab/jordan-model}.

It is worth noting that the test system is appropriate for studies whose conclusions rely on system-level structure and aggregate behavior, such as time-series dispatch and unit commitment analysis, annual energy and resource-adequacy assessment. It is also well suited for renewable integration and curtailment studies, fuel price and fuel supply scenarios, interconnection and import dependence analysis, steady state network studies (such as power flow, optimal power flow, contingency screening), and the benchmarking of methods that require a national-scale case. On the other hand, it is not suitable for drawing operational conclusions about the actual NEPCO grid, whose real parameters, controls, and contracts are confidential and likely differ from the generic values used in this work. In particular, the model should not be used for plant-level commercial analysis. The model also contains no dynamic data, such as machine constants, exciter or governor models, inverter control representations, or fault parameters. Hence, this model is unsuitable for stability, electromagnetic transient (EMT), or protection studies. It also should not be used for reactive power or voltage control planning, since transformer taps are fixed and the compensation is a calibrated aggregate. Finally, this is a synthetic transmission system, therefore it is not suitable for analyses below the transmission level, as the 33\,kV network and the distribution company systems are aggregated onto the 132\,kV buses.

The remainder of the paper is organized as follows. Section~\ref{sec:system} characterizes the Jordanian power system. Section~\ref{sec:construction} describes the construction and synthesis of the test system, including the bus-identity key and temporal alignment of the topology. Section~\ref{sec:validation} reports the validation results. Section~\ref{sec:discussion} discusses assumptions and limitations. Section~\ref{sec:conclusion} presents the conclusion of this work.
\section{The Jordanian Power System}\label{sec:system}
The Jordanian electricity sector evolved from a vertically integrated power authority into an unbundled structure following the sector reforms of the late 1990s, after which generation and distribution assets were separated and partially privatized. However, NEPCO still maintained ownership, operation, and planning of the transmission system, the single-buyer market function, fuel supply for generation, and the cross-border interconnections~\cite{batarseh2022covid}. The transmission network operates at 400~kV and 132~kV; sub-transmission and distribution at 33~kV and below are handled by the regional distribution companies and are outside the scope of the developed model. Table~\ref{tab:char} presents the published characteristics of the 2018 system.

\begin{table}[!h]
\renewcommand{\arraystretch}{1.25}
\caption{Published Characteristics of the Jordanian Power System, 2018~\cite{batarseh2022covid,nepco2018,jica}}
\label{tab:char}
\centering
\begin{tabular}{@{}lll@{}}
\toprule
\textbf{Quantity} & \textbf{Value} & \textbf{Role} \\
\midrule
Peak load (winter, evening) & 3{,}205 MW & Load calibration \\
Peak load (summer) & 3{,}000 MW & --- \\
Minimum load & $\sim$1{,}290 MW & PCM anchor \\
Available capacity & 5{,}236 MW & Gen.\ calibration \\
Combined-cycle capacity & $\sim$2{,}740 MW & Dominant tech.\ \\
Renewable capacity & $\sim$980 MW & RE representation \\
Natural-gas share & $\sim$88\% & Fuel-mix context \\
Purchased-energy cost & 82.82 fils/kWh & Cost cross-check \\
 & ($\sim$117 USD/MWh) & \\
Transmission loss & 1.97\% (energy) & Loss validation \\
400 kV circuit length & 1{,}164 km & Backbone scope \\
132 kV circuit length & 3{,}636 km & Network scope \\
Egypt tie & 400 kV, 550 MW & Boundary inj.\ \\
Syria tie & 400 kV (idle 2018) & Boundary inj.\ \\
Jericho tie & 132 kV (export) & Boundary inj.\ \\
\bottomrule
\end{tabular}
\end{table}

\subsection{Demand}
The 2018 system peak load reached 3{,}205~MW on a winter evening, with a summer peak of 3{,}000~MW. The recorded minimum load was approximately 1{,}290~MW~\cite{nepco2018}. The demand is dominated by occupancy-driven consumption. In NEPCO's reports, household and public-sector use (reported jointly as domestic and government) makes up 45.1\% of the total, commercial and hotel use adds 15.1\%, and industry takes only 22.1\%~\cite{nepco2018}, in additon to \%15 of the load being attributed to water pumping, and the remaining to agriculture and street lightening. This mix is very different from the systems behind the established test cases. The residential share is about 39\% in the United States (with 26\% industrial)~\cite{eia} and about 28\% residential load in the European Union (with 36\% industrial)~\cite{eurostat}. Jordan therefore has an unusually large occupancy driven load and a small industrial base, which generates sharp morning and evening peaks and strong temperature sensitivity~\cite{batarseh2022covid}. 
The 2018 snapshot is adopted as the calibration year throughout, where the demand values of~\cite{batarseh2022covid} for 2019 and 2020 are used only qualitatively, to characterize the load seasonality and shape.

\subsection{Generation}
The 2018 generation fleet had an available capacity of approximately 5{,}236~MW~\cite{nepco2018}, dominated by gas-fired combined-cycle plants (approximately 2{,}740~MW), with steam units, gas turbines, diesel engines, and roughly 980~MW of wind and solar capacity  (as 2018)~\cite{nepco2018}. This capacity distribution resulted in a natural gas annual energy share at close to 88\% of generation in 2018~\cite{nepco2018}. The published per generation company energy accounts also detail the main production plants and their relative output. The Samra combined-cycle station (SEPGCO) is the largest single producer, which is followed by the independent power producers at Amman East and Qatrana~\cite{nepco2018}. The specific plant-level capacities, siting, and cost parameters adopted in the model are provided in Table~\ref{tab:gen}.

\subsection{Interconnections}
NEPCO maintains ties to Egypt (via a 400~kV submarine cable rated at 550~MW), Syria (400~kV), Palestine/Jericho (132~kV), and Iraq~\cite{nepco2018,jica}. In practice, the 2018 interconnection exchanges were not very active. As Jordan was a small net importer from Egypt  (188\,GWh, about 1\% of annual
purchases), the Syrian tie carried no exchange due to regional conditions, and the Jericho connection carried a small export under a contract signed in 2018~\cite{nepco2018} (88\,GWh, about 0.5\%). The interconnections are represented as priced boundary injections at their 2018 levels.

\subsection{Network and Transmission Losses}
By the end of 2018 the system consisted of 1{,}164~km-circuit of 400~kV line (up from 924~km-circuit in 2014, reflecting the initial Green Corridor reinforcements) and 3{,}636~km of 132~kV circuit, across 62 main substations~\cite{nepco2018}. The reported transmission loss for 2018 was 1.97\%~\cite{nepco2018}, which serves as the quantitative loss-validation target described in Section~\ref{sec:losses}. The master plan study documents standard conductor types, transformer sizes, and named line lengths for the main corridors~\cite{jica}.

\section{Construction of the Synthetic Test System}\label{sec:construction} The construction follows the methodology of synthetic networks and extended test databases~\cite{pena2017extended,birchfield2016grid}. The electrical parameters are computed from documented conductor and transformer data or assigned standard values based on their technology, class and age. Whereas aggregate loads are calibrated to the published system snapshot, and the result is validated against real system statistics. However, this work is different from \cite{pena2017extended,birchfield2016grid} in that, rather than being a synthesized network~\cite{birchfield2016grid} or a borrowed canonical one~\cite{pena2017extended}, the model reconstructs the actual Jordanian transmission topology from a published diagram~\cite{al2023global} intended for optimal PMU device placement and verified by the transmission connections, generators and operating substations outlined in \cite{jica}.

\subsection{Topology Reconstruction}
\begin{figure}[!h]
\centering
\includegraphics[width=\columnwidth, height=10cm]{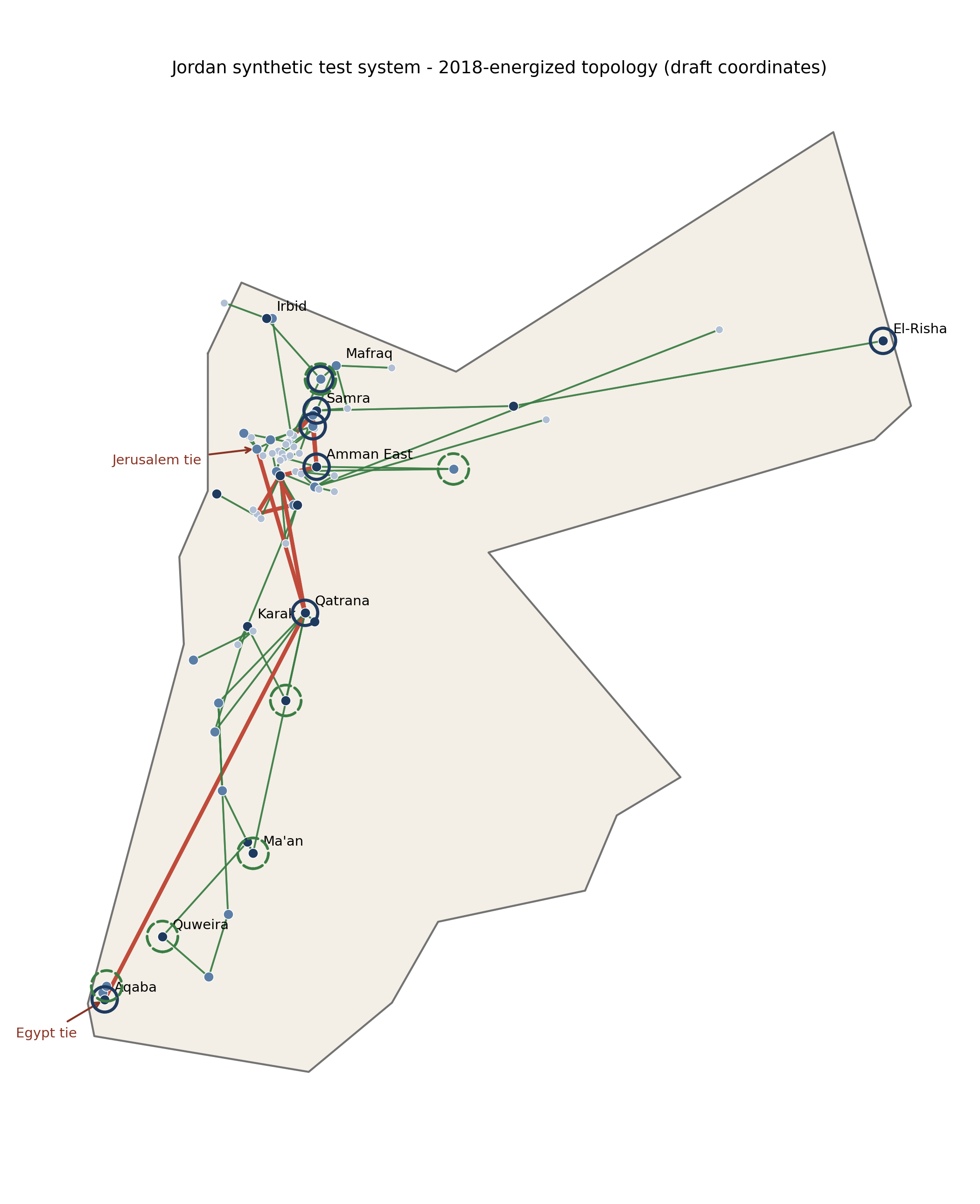}
\caption{The synthetic test system overlaid on a map of Jordan (2018-energized topology): 400~kV backbone (red), 132~kV network (green), thermal plants (dark rings), renewable sites (dashed green rings), and international ties. Node shading indicates the coordinate-confidence tier of the identity key.}
\label{fig:map}
\end{figure}

\begin{figure}[!h]
\centering
\includegraphics[width=0.95\columnwidth]{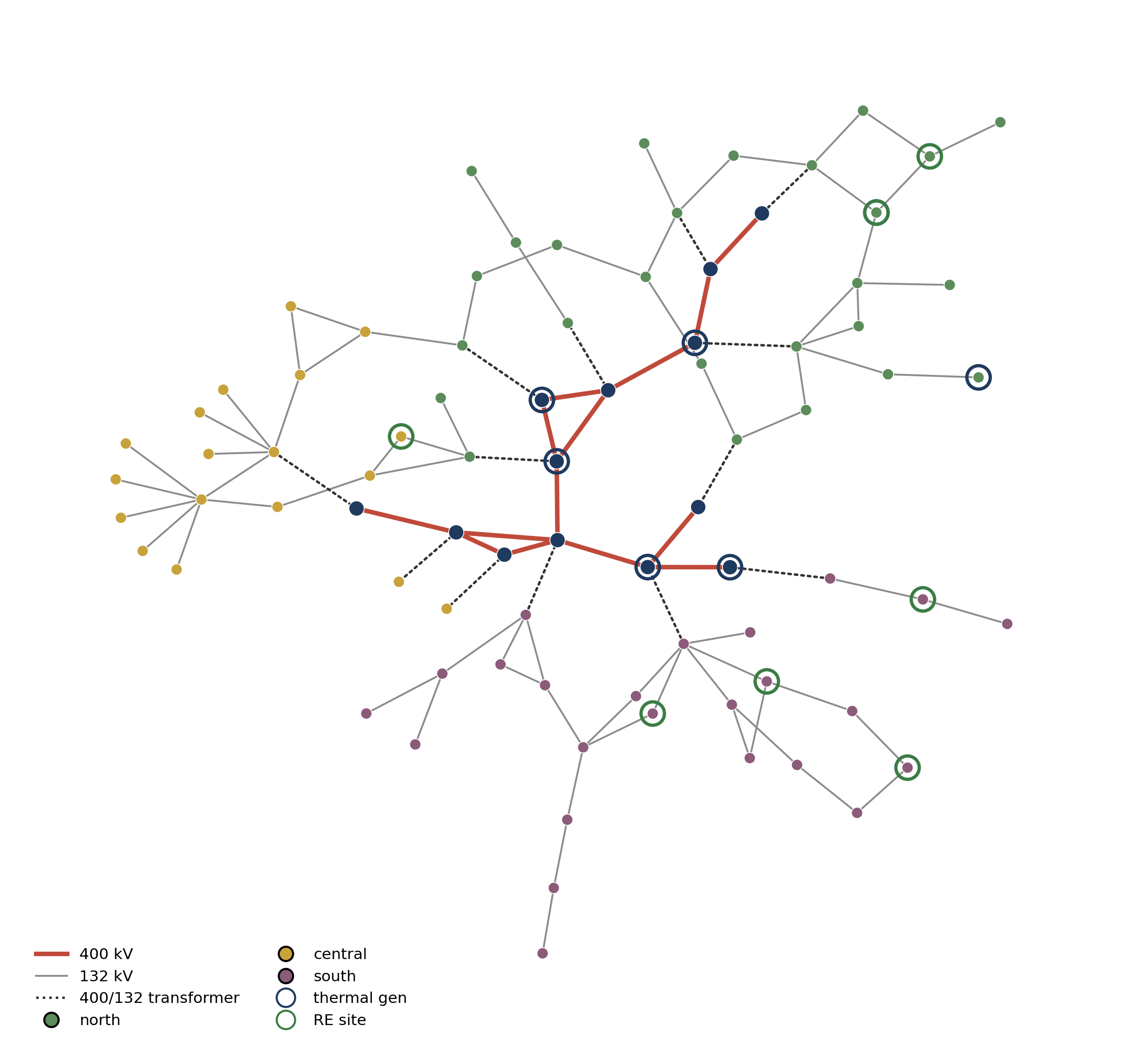}
\caption{Single-line diagram of the synthetic test system. Busbars are colored by subsystem; the 13 backbone substations are each split into a 400~kV busbar (dark) and a 132~kV busbar joined by a transformer (two-circle symbol), expanding the 68 substation busbars to 81 nodes. Red lines are the 400~kV backbone; circles marked G/$\sim$ are conventional/renewable generators.}
\label{fig:sld}
\end{figure}

The transmission topology is reconstructed from the numbered single-line diagram published in~\cite{al2023global}, which illustrates the Jordanian transmission network as 68 numbered buses organized into three regional subsystems corresponding broadly to the northern, central, and southern regions. The diagram was manually parsed region by region to recover the bus-to-bus adjacency, yielding 87 edges spanning all 68 buses with no islanded nodes and a degree distribution consistent with a real transmission graph (average nodal degree 2.47, maximum 8, exponentially decaying tail). This reconstructed edge list serves as the topological structure of the model as shown geographically in Fig.~\ref{fig:map} using the substation coordinates of the bus-identity key (more detail in Section~\ref{sec:key}). The schematic in single-line form, with the 400/132~kV substation splits made explicit, is shown in Fig.~\ref{fig:sld}.

\subsection{Voltage Levels and the 400 kV Backbone}
The single-line diagram represents each substation as a single busbar and does not distinguish between voltage levels. However, in practice thirteen of the 68 substations operated in 2018 as 400/132~kV stations, each containing a 400~kV busbar and a 132~kV busbar joined by transformers. These thirteen are identified by name from the NEPCO master-plan records~\cite{nepco2018,jica} (in the bus-identity key of Section~\ref{sec:key}). Since a power-flow formulation requires one node per busbar, each of the thirteen is represented by two nodes (its original 132~kV bus and a co-located 400~kV bus) joined through a 400/132~kV transformer, which is the same representation used for the multi-voltage transformers of the IEEE 30-bus case, where each transformer connects two existing buses. This expands the 68 substation busbars of the diagram to 81 solved nodes; 68 is the substation count of the source diagram~\cite{al2023global}, while 81 is the node count of the model. The 400~kV buses are interconnected along the documented 400~kV corridors (Sections~\ref{sec:year} and~\ref{sec:lines}). One connection is worth emphasizing, which is the 400~kV line between the Amman East and Amman South substations. Although it is not explicitly identifiable in the source diagram, omitting it separates the 400~kV network into disconnected northern and southern subsystems, preventing the power flow from converging. Therefore, we included this line to ensure network continuity as per the published system maps in the master plan study~\cite{jica}. All validation studies in Section~\ref{sec:validation} are performed on the 81-node model, the only exception is the production cost model, which is solved as a system-level energy-balance dispatch in copper plate mode (its network feasibility is verified after in Section~\ref{sec:pcm}).

\subsection{Topology Reference Year}\label{sec:year}
The topology figure in \cite{al2023global} includes the Green Corridor configuration---the New Ma'an 400/132\,kV substation looped into the Aqaba--Qatrana corridor---which NEPCO's own project tables date to 2019~\cite{nepco2018}. However, because the test case is calibrated to 2018, the network is restored to its 2018-energized state. The New Ma'an loop is removed, the direct Aqaba--Qatrana 400\,kV corridor is restored, Ma'an is represented as the 132\,kV substation it then was in 2018, and the Qatrana--Amman West 400\,kV connection (energized Q4/2018, 110~km-circuit~\cite{nepco2018}) is added even though it is absent in \cite{al2023global}. The restored direct Aqaba--Qatrana corridor is visible in Fig.~\ref{fig:map}. With these dated corrections the model's 400\,kV network totals 1{,}114~km-circuit against NEPCO's published 1{,}164. Where the gap corresponds to the Green-Corridor completed ahead of its 2019 energization. The post-2019 configuration (with the Green-Corridor) is provided with the dataset as a documented scenario variant, separate from the validated 2018 case in \ref{sec:validation}.

\subsection{Line Electrical Parameters}\label{sec:lines}
The line parameters are computed from the documented conductor classes~\cite{jica} using standard overhead-line theory, so that the procedure is fully reproducible. Two conductor classes are used, where the 132~kV lines employ ACSR ``Zebra'' (single conductor per phase), and 400~kV lines employ a twin bundle of large ACSR sub-conductors.

The positive-sequence inductance per unit length is given by:
\begin{equation}
L = 2\times10^{-4}\,\ln\!\left(\frac{\mathrm{GMD}}{\mathrm{GMR}}\right)\quad[\mathrm{H/km}],
\end{equation}
where GMD is the geometric mean distance between phase positions and GMR is the geometric mean radius of the phase conductor (the equivalent bundle GMR for a bundle). The series reactance per unit length at frequency $f$ is:
\begin{equation}
x = 2\pi f\,L = 4\pi f\times10^{-4}\,\ln\!\left(\frac{\mathrm{GMD}}{\mathrm{GMR}}\right)\quad[\Omega/\mathrm{km}],
\end{equation}
which at $f=50$~Hz gives $x = 0.0628\,\ln(\mathrm{GMD}/\mathrm{GMR})$~$\Omega$/km. The positive-sequence shunt capacitance and susceptance per unit length are:
\begin{equation}
C = \frac{2\pi\varepsilon_0}{\ln(\mathrm{GMD}/r_{\mathrm{eq}})},\qquad b = 2\pi f\,C,
\end{equation}
where $r_{\mathrm{eq}}$ is the conductor's physical (or equivalent bundle) radius. The per-phase resistance is the conductor AC resistance at operating temperature divided by the number of sub-conductors $n_b$ in parallel,
\begin{equation}
r = r_{\mathrm{ac}}(T)/n_b\quad[\Omega/\mathrm{km}].
\end{equation}
For the 400~kV twin bundle ($n_b=2$) with sub-conductor GMR $g_c$, physical radius $r_c$, and bundle spacing $d$,
\begin{equation}
\mathrm{GMR}_{\mathrm{bundle}} = \sqrt{g_c\,d},\qquad r_{\mathrm{eq}} = \sqrt{r_c\,d}.
\end{equation}
The geometric assumptions and resulting per-kilometre single-circuit constants are provided in Table~\ref{tab:line}.

\begin{table}[!h]
\renewcommand{\arraystretch}{1.25}
\caption{Conductor Geometry and Derived Per-km Single-Circuit Line Constants (50~Hz)}
\label{tab:line}
\centering
\begin{tabular}{@{}lcc@{}}
\toprule
\textbf{Parameter} & \textbf{132 kV} & \textbf{400 kV} \\
 & (Zebra, single) & (2$\times$ACSR) \\
\midrule
Sub-conductors $n_b$ & 1 & 2 \\
Bundle spacing $d$ & --- & 0.45 m \\
Conductor GMR $g_c$ & 0.0114 m & 0.0127 m \\
Conductor radius $r_c$ & 0.0143 m & 0.0158 m \\
Phase spacing GMD & $\sim$5 m & $\sim$11 m \\
Resistance $r$ & 0.075 $\Omega$/km & 0.025 $\Omega$/km \\
Reactance $x$ & 0.40 $\Omega$/km & 0.30 $\Omega$/km \\
Capacitance $c$ & 8.9 nF/km & 12.75 nF/km \\
Rating (1 circuit) & $\sim$145 MVA & $\sim$1{,}000 MVA \\
\bottomrule
\end{tabular}
\end{table}

The master-plan study states that double-circuit construction is the transmission standard~\cite{jica}, where each corridor is modeled as two identical parallel circuits. Neglecting inter-circuit coupling,
\begin{equation}
r_{\mathrm{dc}}=\tfrac{r}{2},\quad x_{\mathrm{dc}}=\tfrac{x}{2},\quad c_{\mathrm{dc}}=2c,\quad \mathrm{rating}_{\mathrm{dc}}=2\times\mathrm{rating}.
\end{equation}
For the power-flow and OPF studies the double-circuit equivalent is used as a single aggregated branch. For the $N\!-\!1$ study the two circuits are represented explicitly as parallel single circuits. All branch quantities are converted to per unit on a 100~MVA base, with $Z_{\mathrm{base}}=V_{\mathrm{base}}^2/S_{\mathrm{base}}$ (174.24~$\Omega$ at 132~kV, 1{,}600~$\Omega$ at 400~kV), so that for a line of length $\ell$
\begin{equation}
r_{\mathrm{pu}}=\frac{r\,\ell}{Z_{\mathrm{base}}},\quad x_{\mathrm{pu}}=\frac{x\,\ell}{Z_{\mathrm{base}}},\quad b_{\mathrm{pu}}=(2\pi f\,c\,\ell)\,Z_{\mathrm{base}}.
\end{equation}

Assigning lengths to each line presented a challenge, as public sources provide incomplete data. NEPCO and master-plan project tables state route lengths for eighteen individual lines~\cite{nepco2018,jica}, while NEPCO additionally publishes the system's total circuit-kilometres by voltage level and year~\cite{nepco2018}. To resolve this, line lengths are assigned using a three-tiered approach.
First, where a line is individually named in the project tables, such as the main southern 400~kV routes and the Qatrana--Amman West connection—its published length is used directly. Second, for the 400~kV segments of the Amman ring that lack published lengths, map-estimated values are used. Third, the remaining 132~kV lines are assigned typical route lengths based on regional substation spacing; 15~km in the dense north and north-center, 25~km in the center, and 45~km in the sparse south. While these 132~kV lengths are geographic assumptions and not from official NEPCO values, the resulting total aligns with NEPCO’s published 3{,}636~km-circuit to within $+5\%$ (Table~\ref{tab:struct}).
Finally, a uniform conductor class per voltage level is used, which the loss analysis in Section~\ref{sec:losses} identifies as a residual source of loss underestimation compared to the annual loss by NEPCO.

\subsection{Transformers}
The 400/132~kV transformers are represented with the standard ratings (paired 400~MVA units, consistent with documented standard sizes and the 2$\times$400~MVA installation at Amman~West commissioned in 2018~\cite{nepco2018}) and typical short-circuit impedance. The 132/33~kV transformation and the underlying distribution systems are not modeled; the 33~kV and bulk-supply-point load is aggregated onto their corresponding 132~kV buses.

\subsection{Load Representation}\label{sec:load}
The spatial load model follows the participation-factor approach used in ~\cite{pena2017extended}. With $\hat{P}_i$ the measured bulk-supply-point peak at bus $i$~\cite{jica}, the bus active loads are obtained by scaling the measured peaks to the calibrated system peak $P_{\mathrm{sys}}=3{,}205$~MW:
\begin{equation}
P_i = \hat{P}_i\,\frac{P_{\mathrm{sys}}}{\sum_j \hat{P}_j}.
\end{equation}
The measured peaks sum to 2{,}928~MW, giving a scaling factor $3{,}205/2{,}928 = 1.095$. The load participation factor is:
\begin{equation}
\alpha_i = \frac{P_i}{\sum_j P_j},\qquad \sum_i \alpha_i = 1,
\end{equation}
reported by region in Table~\ref{tab:load}. The reactive demand uses a lagging power factor of 0.88 \cite{nepco2018},
\begin{equation}
Q_i = P_i\,\tan\phi,\qquad \tan\phi = \tan(\cos^{-1}0.88) = 0.540,
\end{equation}
and shunt capacitor compensation is placed at the larger load buses,
\begin{equation}
B_{\mathrm{sh},i} = \kappa\,Q_i,\quad \kappa = 0.75,\ \text{where } P_i>15~\mathrm{MW}.
\end{equation}
Of the 68 buses, 51 are loads buses and 17 are pure transhipment (zero injection bus) or generation nodes, where the northern region contains roughly half of the system load.

The synthesized hourly demand series is calibrated to three NEPCO report published values; the 2018 peak of 3{,}205\,MW, the annual purchased energy of 18{,}913\,GWh, and the implied load factor of 0.674. The temperature-sensitivity coefficient is bisected automatically to the load-factor target (calibrated $K_{\mathrm{cool}}=0.0269$ per cooling degree-hour). Validation is performed on the dispatch outputs, not on demand.

\begin{table}[!h]
\renewcommand{\arraystretch}{1.25}
\caption{Load Distribution and Participation Factors by Subsystem}
\label{tab:load}
\centering
\begin{tabular}{@{}lcccc@{}}
\toprule
\textbf{Subsystem} & \textbf{Load buses} & \textbf{Load (MW)} & \textbf{$\sum\alpha$} & \textbf{$\alpha$ range} \\
\midrule
North & 21 & 1{,}703 & 0.531 & 0.003--0.059 \\
Central & 12 & 821 & 0.256 & 0.001--0.048 \\
South & 18 & 681 & 0.212 & 0.001--0.028 \\
\textbf{Total} & \textbf{51} & \textbf{3{,}205} & \textbf{1.000} & --- \\
\bottomrule
\end{tabular}
\end{table}

\subsection{Generation Representation}
Generators are placed on their host substations using the bus-identity key of Section~\ref{sec:key} and sized to documented capacities~\cite{nepco2018}. The Samra station is assumed to be the system slack bus. The three independent producers of the Al~Manakher complex---Amman East (IPP1), Amman Asia (IPP3), and Levant (IPP4)---share the Amman East 400~kV substation. The relevant technical data for the generators are provided in Table~\ref{tab:gen}.

\subsection{Generator Cost Curves}\label{sec:gencost}
Each thermal unit is given a quadratic cost curve derived from its heat rate, following the standard input--output formulation~\cite{woodwollenberg}. The fuel input is modeled as $I(P)=a+bP+cP^2$ (MMBtu/h), so that the heat rate $\mathrm{HR}(P)=1000\,I(P)/P$ and the cost $C(P)=\pi_f\,I(P)=c_2P^2+c_1P+c_0$, where $\pi_f$ is the fuel price. The three coefficients are fixed by three conditions: the full-load heat rate equals a representative value $\mathrm{HR}_{\mathrm{full}}$ for the unit's technology and commissioning year, the part load heat rate at minimum stable output is $\mathrm{HR}_{\mathrm{full}}(1+K)$ with $K=0.12$ (a typical 12\% part-load degradation); and the heat rate is minimized at full output ($c=a/P_{\max}^2$).

Full-load heat rates are assigned by technology and commissioning year from representative literature values, consistent with the reported heat rates by prime mover in~\cite{eia}. Combined-cycle units are assigned 6{,}400--6{,}700~Btu/kWh, with newer plants at the lower end. The reciprocating-engine plants are assigned 7{,}500--8{,}000~Btu/kWh. The figure for the IPP3 engines follows the manufacturer specification~\cite{wartsila}, and IPP4's medium-speed engines are assigned 7{,}500~Btu/kWh. Legacy gas turbines are assigned 11{,}000--11{,}500~Btu/kWh and conventional steam 10{,}200~Btu/kWh.

Fuel prices reflect the 2018 Jordanian supply. Imported natural gas is priced at 8~USD/MMBtu, Egyptian pipeline gas near 5~USD/MMBtu and Brent-indexed LNG near 9~USD/MMBtu. Heavy fuel oil for the Aqaba steam plant is priced at 12~USD/MMBtu. Domestic field gas for the Risha plant is priced at 2.0~USD/MMBtu, corresponding to the fixed rate of 50~fils (subvalue of dinar)/m$^3$ reported in the operator's documents~\cite{cegco}.

The resulting full-load marginal costs are listed in Table~\ref{tab:gen}. Risha is lowest on cheap domestic gas, the imported-gas combined-cycle fleet follows at 51--54~USD/MWh, and the HFO-fired steam plant is the most costly. These marginal fuel costs are distinct from and lower than the published all-in purchased-energy cost of 82.82~fils/kWh which yields to approximately 117~USD/MWh~\cite{nepco2018} as this cost includes capacity payments.

\begin{table}[!h]
\renewcommand{\arraystretch}{1.2}
\caption{Generator Data (2018 Calibration). MC is the Full-Load Marginal Cost of the Heat-Rate Curve (Section~\ref{sec:gencost}); Bus 2$xx$ is the 400~kV Node Co-located with Bus $xx$. $^\dagger$Risha's 80~MW is the Instantaneous Fuel Deliverability; its Annual Cap is Tighter (Section~\ref{sec:pcm})}
\label{tab:gen}
\centering
\begin{tabular}{@{}llccll@{}}
\toprule
\textbf{Plant} & \textbf{Bus} & \textbf{MW} & \textbf{Tech.} & \textbf{Fuel} & \textbf{mc} \\
 & & & & & \textbf{(\$/MWh)} \\
\midrule
Samra (SEPGCO) & 211 & 1{,}150 & CCGT & gas & 53.6 \\
Amman East (IPP1) & 230 & 380 & CCGT & gas & 53.2 \\
IPP3 (Amman Asia) & 230 & 573 & ICE & gas & 64.0 \\
IPP4 (Levant) & 230 & 250 & ICE & gas & 60.0 \\
Qatrana (IPP2) & 257 & 373 & CCGT & gas & 52.8 \\
Zarqa (ACWA) & 229 & 485 & CCGT & gas & 51.2 \\
Aqaba steam (CEGCO) & 261 & 400 & Steam & HFO & 122.4 \\
Rehab (CEGCO) & 12 & 297 & CCGT & gas & 92.0 \\
Risha (CEGCO) & 21 & 80$^\dagger$ & GT & gas$^\dagger$ & 22.0 \\
Solar fleet (5 sites) & 63,64,12,60,20 & 447.5 & Solar & --- & 0 \\
Wind fleet (3 sites) & 51,63,6 & 280.4 & Wind & --- & 0 \\
Egypt import & 261 & 0--120 & Tie & --- & 117 \\
\bottomrule
\end{tabular}
\end{table}

\subsubsection{Renewable Fleet and Profiles}\label{sec:re}
The 2018 renewable fleet is represented plant-by-plant. Table~\ref{tab:fleet} lists the solar and wind inventory, where capacities are the AC/grid-connected values from NEPCO's project and capacity tables~\cite{nepco2018}, and each plant is assigned its commercial operation date, so that the model applies a time-varying availability. The utility fleet totals 447.5,MW of solar and 280.4,MW of wind, exactly reproducing NEPCO's transmission-connected capacity accounts. Distributed net-metering and wheeling capacity (250,MW) is excluded. The validation values for this model relies on NEPCO's purchases, which do not account for the distributed generation. Instead, the effect of this distributed capacity is already implicitly captured within the net demand time series.

\begin{table}[!h]
\caption{2018 Utility-Scale Renewable Fleet}
\label{tab:fleet}
\centering
\begin{tabular}{lccc}
\hline
Plant & MW$_{\mathrm{ac}}$ & In service & Host bus \\
\hline
Tafila WF          & 117.0 & Sep.\ 2015 & 51 (Hasa) \\
Ma'an (Al-Hussein) WF & 80.0 & Q1 2016 & 63 \\
Al Rajef WF        & 82.0  & Apr.\ 2018$^{\dagger}$ & 63 \\
Hofa pilot         & 1.4   & legacy & 6 \\
Shams Ma'an        & 52.5  & Oct.\ 2016 & 63 \\
Round-1 Ma'an cluster & 90.0 & 2016 & 63 \\
Round-1 other sites   & 34.5 & 2016 & 60, 12, 20 \\
Round-1 unattributed  & 75.5 & 2016 & 63$^{\ddagger}$ \\
Quweira (Sheikh Zayed) & 95.0 & Apr.\ 2018 & 64 \\
Mafraq I + Empire (FRV) & 100.0 & Jun.\ 2018 & 12 \\
\hline
\end{tabular}
\\[2pt]\footnotesize $^{\dagger}$pre-COD export; COD Oct.\ 2018.\quad
$^{\ddagger}$Round-1 capacity not yet attributed to named PPAs;
placed in the Ma'an development area.
\end{table}

Hourly profiles are generated per plant. Solar irradiance is taken from the NSRDB Meteosat Prime Meridian satellite at each site's coordinates and converted with a fixed-tilt PVWatts (14\% system losses, DC/AC ratio 1.2 with inverter clipping), where fixed-tilt mounting reflects the pre-dominant configuration of Jordan's Round-1 fleet. Wind capacity factors are taken from MERRA-2 reanalysis~\cite{gelaro2017} via Renewables.ninja~\cite{staffell2016}, with a uniform turbine representation (V112-class, 84\,m hub height). Because reanalysis is known to underestimate the production of the sites, a single fleet-level bias factor of 1.226 is derived from the 2016--2017 observed-to-modeled energy ratio~\cite{nepco2018} following the bias correction approach of~\cite{staffell2016}.

The final output is capped at the plant nameplate capacity. This saturation only happens during the highest wind speed hours. Users of these time series should note two key consequences of our data adjustments. First, hourly wind outputs are scaled by a factor of 1.226, which amplifies wind ramp rates relative to the raw reanalysis data. Also, because reanalysis data inherently smooths subgrid variability, the net error is ambiguous, meaning these wind ramp statistics should be treated as approximate ramps.

\subsection{Bus--Substation Identity Key}\label{sec:key}
The PMU study from which the topology is reconstructed~\cite{al2023global} does not name its buses. We therefore build the bus--substation identification by triangulating four sources. Namely, the named substations and lines of the NEPCO system map, project tables~\cite{nepco2018,jica}, the regional bulk-supply-point lists of the distribution companies, the network's own structure (node degrees, chain patterns, voltage levels), and a small set of unambiguous spatial anchors (the Egypt interconnection at Aqaba, the Jerusalem tie at Amman West, the Syria tie at Amman North, and the isolated eastern radial terminating at the Risha gas field), in addition to the author's own knowledge of the Jordanian regions. Bus assignments propagate outward from these spatial anchors and are labeled in the data with a three-tier confidence label; high where an assignment is forced by an anchor, a named line length (17 buses); medium where structure and geography agree but no single source is decisive (19 buses); and low for the remaining metropolitan 132\,kV buses whose disambiguation (e.g., among adjacent Amman substations) does not affect any result in this paper (32 buses). The full key, with the evidence string for every bus, is distributed with the dataset. Fig.~\ref{fig:map} illustrates it geographically.

\section{Validation}\label{sec:validation}
The model was validated with the open-source pandapower package~\cite{thurner2018pandapower} and is also released in MATPOWER format~\cite{zimmerman2010matpower}, with the cross-solver agreement demonstrated in Section~\ref{sec:cross}. Table \ref{tab:summary} summarizes the studies and their outcomes.

\begin{table}[!h]
\renewcommand{\arraystretch}{1.25}
\caption{Test System Dimensions and Validation Results}
\label{tab:summary}
\centering
\begin{tabular}{@{}ll@{}}
\toprule
\textbf{Quantity} & \textbf{Value} \\
\midrule
Buses (132 kV + 400 kV) & 81 \\
Line circuits / transformers & 170 (85 corridors) / 26 (13 subst.) \\
Calibration year & 2018 \\
System peak load & 3{,}205 MW (winter) \\
AC power flow & converges; $V$ 0.951--1.033 pu;  loss 1.27\% \\
AC-OPF / DC-OPF & 131{,}478 / 129{,}288 \$/h ($-1.7\%$) \\
$N\!-\!1$ (196 cont.) & 94\% secure; 0 islanding \\
Annual energy mix & top-3 CCGT $\le+3.8\%$; solar $-2.0\%$; wind $-21\%$ \\
Annual losses & 0.70\% series; 1.2--1.5\% comparable vs 1.97\% \\
Structural & $m/n$ 1.21 vs 1.22 quota; ckt-km $+5\%$ / $-4\%$ \\
Cross-solver & MATPOWER $\Delta V \le 4.9\times10^{-15}$ pu \\
\bottomrule
\end{tabular}
\end{table}

\subsection{Structural Statistics}\label{sec:structural}
Table~\ref{tab:struct} evaluates the network against the structural criteria proposed for synthetic grids in~\cite{birchfield2016grid}, computed at the corridor level (where parallel lines are collapsed into one). 
Since part of the motivation is that the Jordanian system differs from the the established benchmarks, it is worth commenting about the use of criteria derived from North American systems to evaluate a Middle-Eastern network~\cite{birchfield2016grid}. However, the characteristics that differentiate the Jordanian system are operational, such as its fuel-supply structure, demand composition, and exposure to external disruption. The criteria of Table~\ref{tab:struct}~\cite{birchfield2016grid} are structural, as they describe properties of transmission networks as engineered graphs, which arise from engineering economics common to all such systems and have been found to hold across networks on several continents~\cite{pagani2013}. Structural agreement with real-grid ranges is therefore evidence that the reconstruction generated a plausible transmission network, without implying operational similarity.

After the structural validation, the lines-per-node ratio of 1.21 sits at the real-grid quota of 1.22 used in~\cite{birchfield2016grid}, where 19\% of substations carry a higher-voltage bus, which is inside the reported 10--20\% range, the substation composition (76\% load-only, 13\% with generation) matches the proportions reported for the Eastern Interconnect, and the degree distribution decays exponentially with a maximum degree of~7. Two aggregate comparisons tie the synthetic parameters to published system totals, where modeled 132\,kV circuit-kilometres agree with NEPCO's 3{,}636\,km to within $+5\%$, and the 400\,kV network agrees as described in Section~\ref{sec:year}. The graph diameter of 14 exceeds the $\sqrt{N}\!\approx\!9$ value typical of compact, mesh-like interconnections. This deviation has a direct geographic explanation, in which the diameter is mostly realized between the Waqas and Ghor Safi substations, the northern and southern extremities of the Jordan Valley, which are connected through radial lines spanning nearly the full length of the country. A network of this shape shows a larger diameter than a compact mesh of equal size, and the deviation is therefore attributed to the geography.

\begin{table}[!h]
\caption{Structural Statistics vs.\ Real-Grid Criteria}
\label{tab:struct}
\centering
\begin{tabular}{lcc}
\hline
Statistic & Model & Reference~\cite{birchfield2016grid} \\
\hline
Lines per node ($m/n$)        & 1.21  & 1.22 quota \\
Avg.\ node degree             & 2.42  & 2.2--3.0 \\
Max.\ node degree             & 7     & single-digit \\
Degree-1 bus share            & 26\%  & exponential tail \\
Dual-voltage substations      & 19\%  & 10--20\% \\
Load-only substations         & 76\%  & $\approx$80\% (EI) \\
Substations with generation   & 13\%  & $\approx$10.5\% (EI) \\
Avg.\ clustering coefficient  & 0.089 & 0.05--0.15 \\
Connected components          & 1     & 1 \\
132\,kV circuit-km            & 3{,}822 & 3{,}636 (NEPCO) \\
400\,kV circuit-km            & 1{,}114 & 1{,}164 (NEPCO)$^{\dagger}$ \\
\hline
\end{tabular}

\end{table}

\subsection{Power Flow}
\begin{figure}[!h]
\centering
\includegraphics[width=0.9\columnwidth]{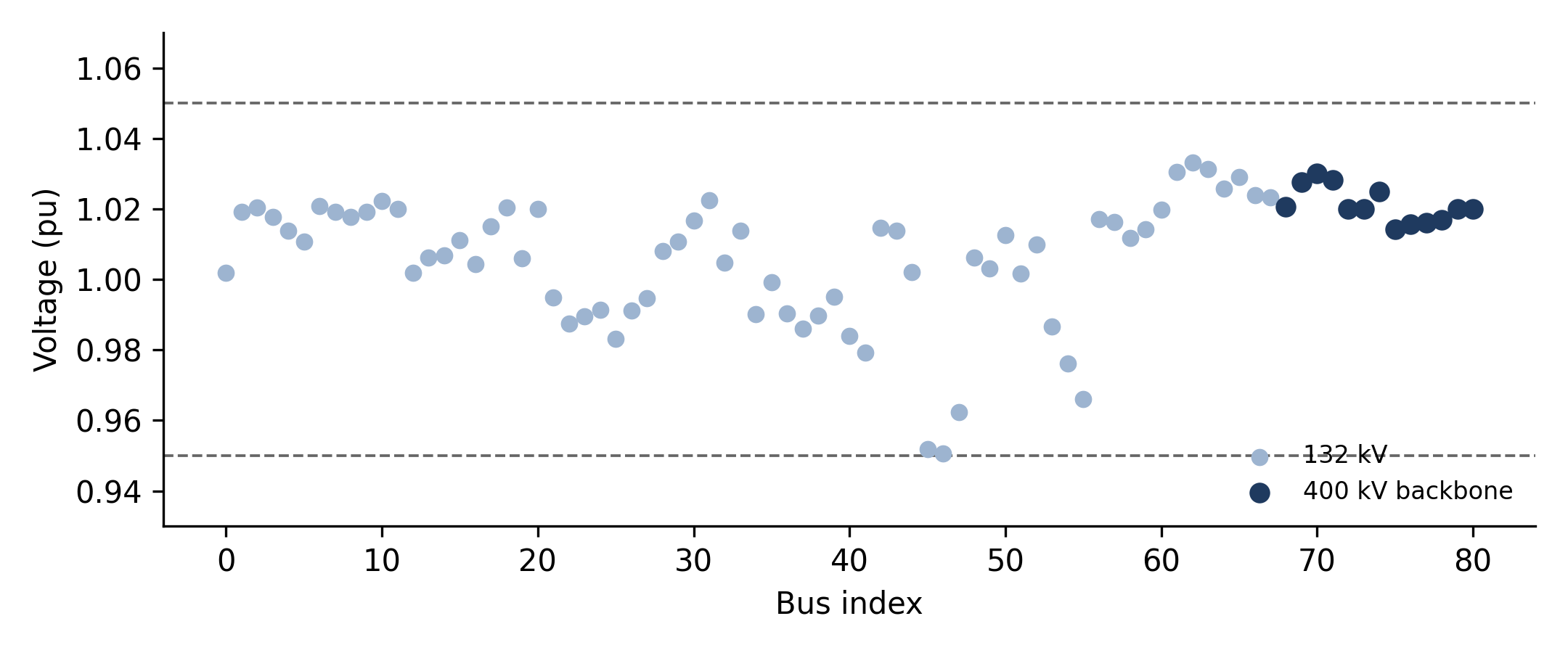}
\caption{AC power-flow bus voltage profile at the 2018 peak. Dark markers: 400~kV backbone buses; light markers: 132~kV buses.}
\label{fig:volt}
\end{figure}

\begin{figure}[!h]
\centering
\includegraphics[width=0.9\columnwidth]{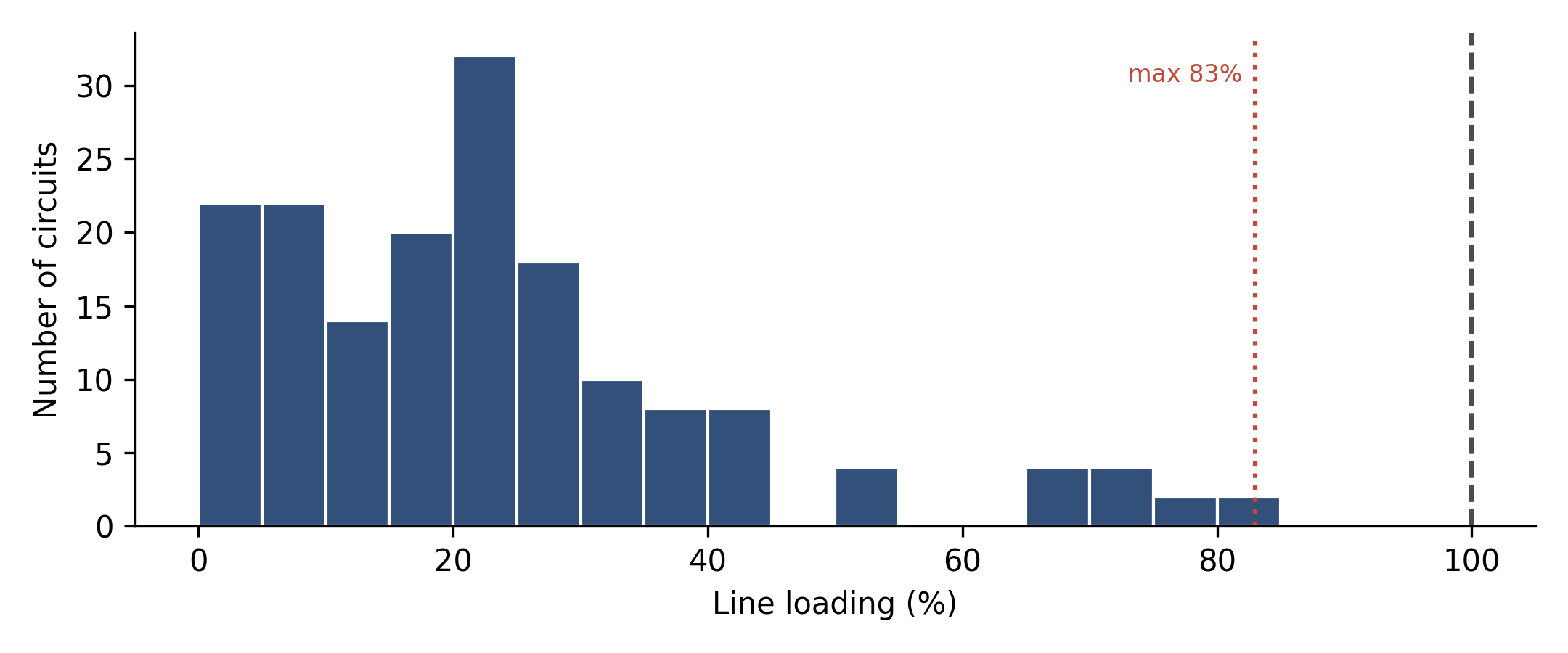}
\caption{Distribution of line loading at the 2018 peak AC power flow. No line exceeds 100\% (dashed); the maximum is 83\%.}
\label{fig:load}
\end{figure}

The AC power flow converges with a clean voltage profile. All bus voltages lie within 0.951--1.033~pu with no violations, and the most heavily loaded line reaches 83\% of its rating. Total series losses at the 2018 peak are 40.91\,MW (1.27\% of load). We however emphasize that a single-snapshot loss percentage is not comparable to NEPCO's annual 1.97\% energy figure; the energy-weighted comparison is made in Section~\ref{sec:losses}. Figs.~\ref{fig:volt} and~\ref{fig:load} show the voltage profile and line-loading distribution.

\subsection{Optimal Power Flow}
The AC optimal power flow converges to \$131{,}478/h at the short-run marginal costs of Table~\ref{tab:gen} (voltages 0.992--1.050~pu, maximum loading 82\%), and the DC OPF to \$129{,}288/h ($-1.7\%$). The dispatch reproduces the system's merit order, where renewables are dispatched in full, Risha loads fully on cheap domestic gas, the combined-cycle fleet carries the bulk of demand with the marginal CCGT setting a system price of $\approx$\$52/MWh, and the engine, steam, and import resources remain at a minimum. 
At this operating point the dispatch cost per unit of demand is $\approx$\$45/MWh, which sits below the published average purchase cost of $\approx$\$117/MWh~\cite{nepco2018}, as the latter is an annual all inclusive figure with capacity payments and PPA terms that a dispatch cost excludes.
Therefore the economic ordering converges and emerges independently from heat-rate-derived costs.

\subsection{$N\!-\!1$ Contingency Screening}\label{sec:n1}
From the base case (DC-OPF dispatch, 84\% maximum base loading), all 196 single-circuit and single-transformer outages are screened with DC power flow. The system is $N\!-\!1$ secure for 94\% of contingencies, with no islanding. Twelve outages produce post-contingency over loading, listed by substation in Table~\ref{tab:n1}. The pattern is physically interpretable, where the binding contingencies are the metropolitan 132\,kV corridors serving the Amman and Zarqa load pockets (highest population densities in Jordan), and the Qatrana--Ma'an circuit that connects the southern renewable generation cluster.

\begin{table}[!h]
\caption{Overloading $N\!-\!1$ Contingencies (one circuit of)}
\label{tab:n1}
\centering
\begin{tabular}{lc}
\hline
Outaged corridor & Post-cont.\ loading \\
\hline
Amman South -- Madaba South (132) & 148\% \\
Zerqa -- New Zarqa (132)          & 140\% \\
Amman North -- Marka (132)        & 130\% \\
Qatrana -- Ma'an RE hub (132)     & 121\% \\
400/132 transformer, Amman South  & 109\% \\
Amman South -- Ashrafia (132)     & 106\% \\
\hline
\end{tabular}
\\[2pt]\footnotesize Each row represents both circuits of the corridor (12 contingencies in total); medium/low-confidence bus names carry their crosswalk labels in the dataset.
\end{table}

\subsection{Time-Series Dispatch (Representative Weeks)}
\begin{figure*}[!h]
\centering
\includegraphics[width=0.9\textwidth]{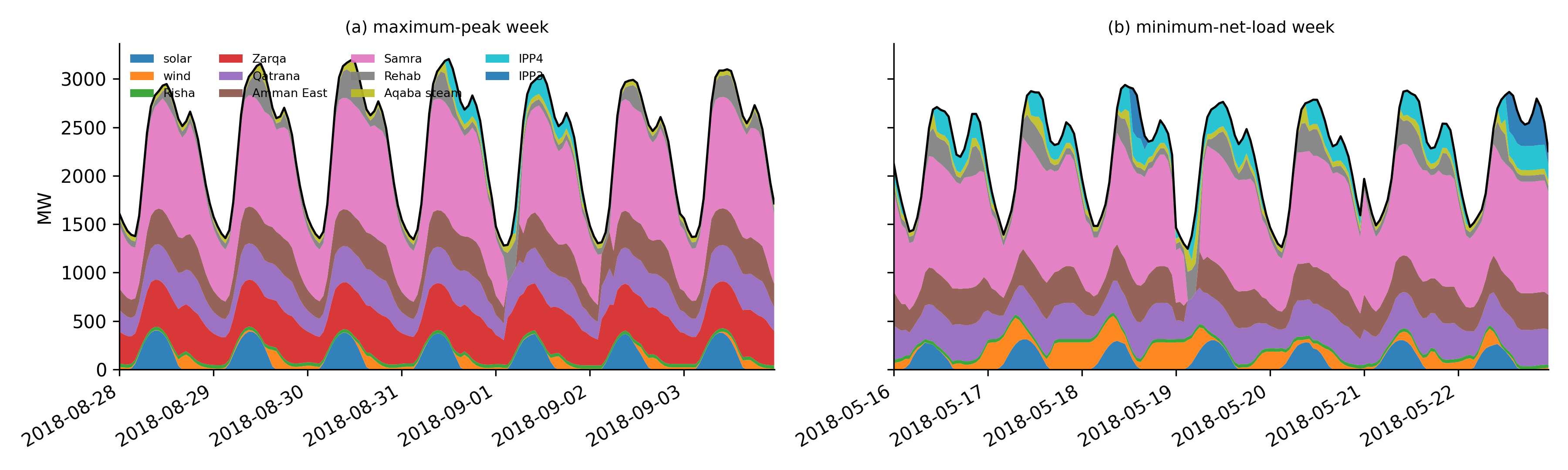}
\caption{Hourly generation dispatch for the two representative weeks (black line: demand). In the winter peak week (a) the combined-cycle fleet cycles and peaking/steam units appear at the daily peaks; in the spring minimum-net-load week (b) solar carves out the midday net load and the peakers remain off.}
\label{fig:disp}
\end{figure*}

To illustrate model behavior, the production cost system is run at hourly resolution for two representative weeks~\cite{pena2017extended}. The first is a winter week including a peak load and low solar output. The second is a spring week with high solar output and minimum net load, which is defined as total demand with renewable generation substracted. The resulting profiles and dispatch stacks appear in Fig.~\ref{fig:disp}. Because actual unit commitment data is private, generator parameters such as minimum stable levels and startup costs are assigned typical technology values~\cite{pena2017extended} as listed in Table~\ref{tab:uc}.
During the winter week, the combined cycle fleet cycles to meet evening peaks with a maximum net load ramp of 342~MW per hour. During the spring week, solar generation significantly reduces midday net load, allowing the most expensive units to remain off, without renewable curtailment (as reflected in the 2018 grid conditionins \cite{nepco2018}. The complete quantitative validation is performed using the full year run detailed in Section~\ref{sec:pcm}.

\begin{table}[!h]
\renewcommand{\arraystretch}{1.2}
\caption{Typical-by-Technology Unit-Commitment Parameters}
\label{tab:uc}
\centering
\begin{tabular}{@{}lcccc@{}}
\toprule
\textbf{Technology} & \textbf{$P_{\min}$ (\%)} & \textbf{Min up (h)} & \textbf{Min down (h)} & \textbf{Start (\$/MW)} \\
\midrule
Combined cycle & 45 & 4 & 3 & 60 \\
Gas turbine & 25 & 1 & 1 & 30 \\
Steam & 40 & 8 & 6 & 100 \\
Wind / solar & 0 & --- & --- & 0 \\
\bottomrule
\end{tabular}
\end{table}

\subsection{Annual Production-Cost Validation}\label{sec:pcm}
\begin{figure}[!h]
\centering
\includegraphics[width=0.95\columnwidth]{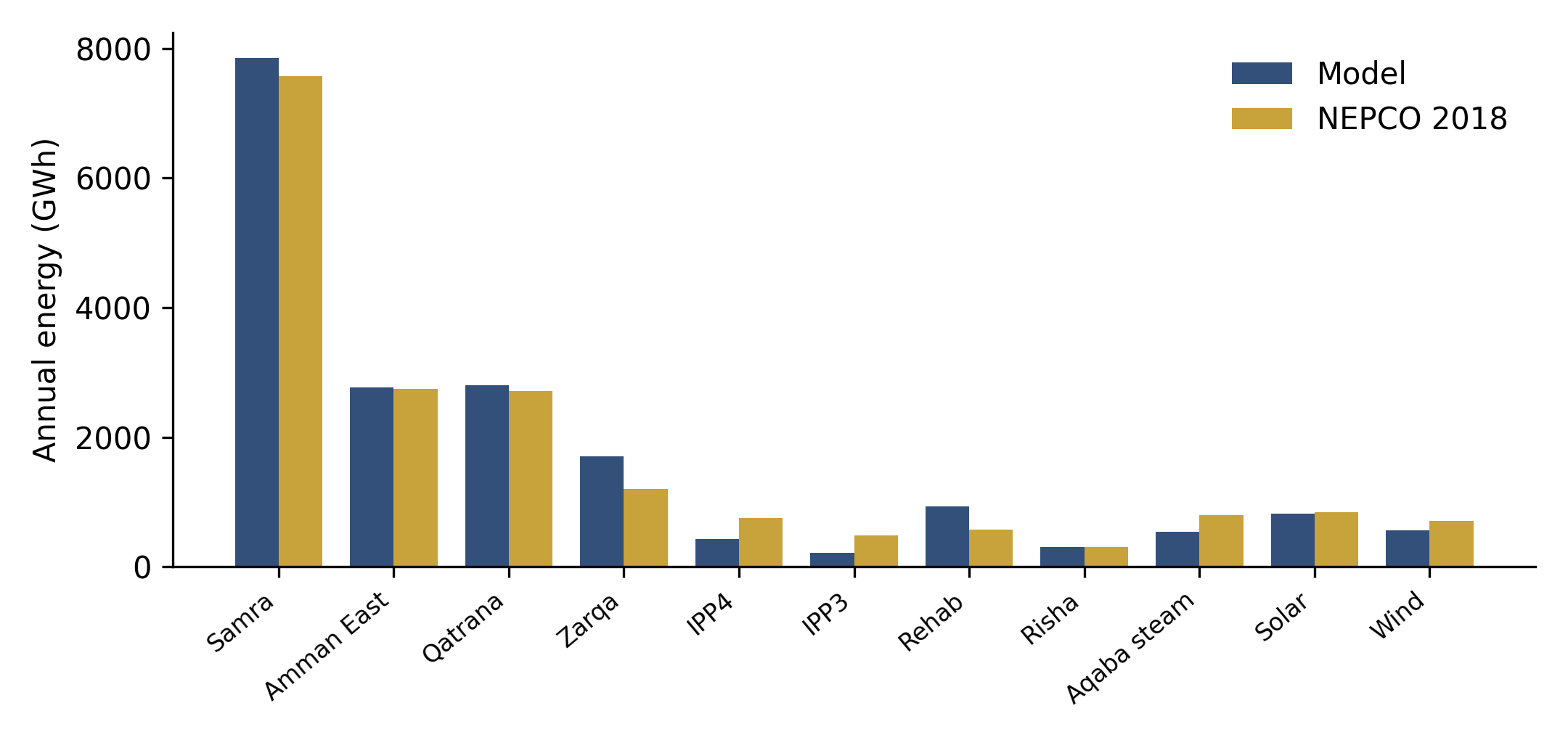}
\caption{Annual generation by plant: model versus NEPCO published 2018.}
\label{fig:mix}
\end{figure}

\begin{figure}[!h]
\centering
\includegraphics[width=0.95\columnwidth]{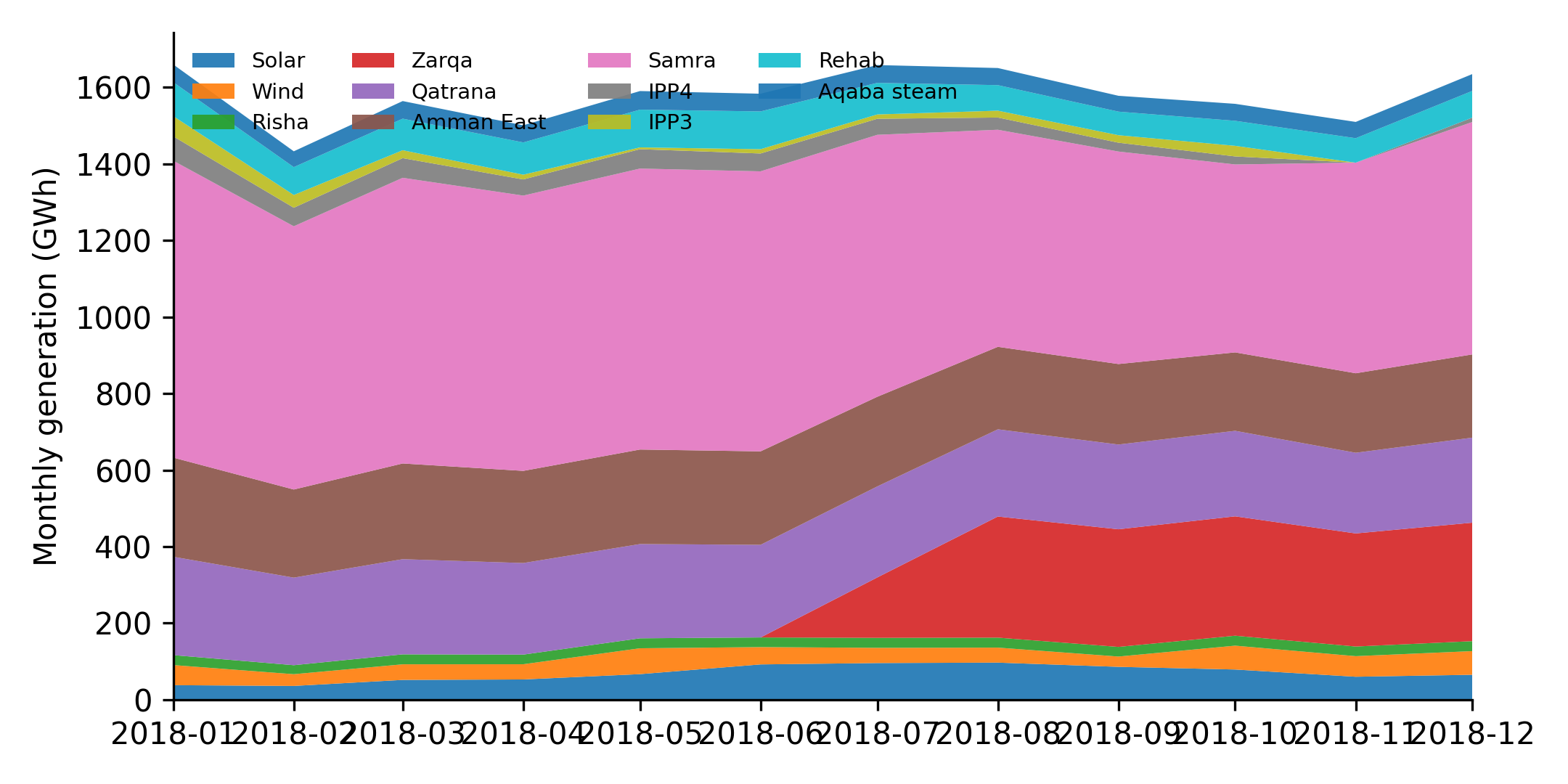}
\caption{Monthly generation by plant over the full 2018 weather year, showing the seasonal dispatch pattern and the summer cooling-driven peak.}
\label{fig:annualstack}
\end{figure}

The production-cost model performs coupled unit commitment and economic dispatch over all 8{,}760 hours of 2018 using the synthesized load, solar, and wind time series. The dispatch minmizes operational costs, given the quadratic heat-rate curves of Section~\ref{sec:gencost} and operational constraints.Table~\ref{tab:mix} compares the modeled annual energy per plant against NEPCO's purchases~\cite{nepco2018}. The comparison is free of parameter tuning as no cost parameter was adjusted to improve agreement. Figs.~\ref{fig:mix} and~\ref{fig:annualstack} visualize the comparison and the monthly dispatch.

\begin{table}[!h]
\caption{Modeled vs.\ Reported Annual Energy, 2018 (GWh)}
\label{tab:mix}
\centering
\begin{tabular}{lrrr}
\hline
Plant & Model & NEPCO & Dev. \\
\hline
Samra (SEPGCO)      & 7{,}852 & 7{,}568 & $+3.8\%$ \\
Amman East (IPP1)   & 2{,}763 & 2{,}740 & $+0.8\%$ \\
Qatrana (IPP2)      & 2{,}806 & 2{,}713 & $+3.4\%$ \\
Zarqa (ACWA)        & 1{,}699 & 1{,}198 & $+42\%$  \\
IPP4 (Levant)       &    432  &    752  & $-43\%$  \\
IPP3 (Amman Asia)   &    212  &    486  & $-56\%$  \\
Rehab               &    936  &    576  & $+62\%$  \\
Risha               &    307  &    305  & $+0.7\%$ \\
Aqaba steam         &    537  &    792  & $-32\%$  \\
Flexible segment$^{\ast}$ & 3{,}816 & 3{,}804 & $+0.3\%$ \\
\midrule
Solar               &    819  &    836  & $-2.0\%$ \\
Wind                &    560  &    705  & $-21\%$  \\
\hline
Total               & 18{,}923 & 18{,}913 & $+0.1\%$ \\
\hline
\end{tabular}
\\[2pt]\footnotesize $^{\ast}$Zarqa + IPP4 + IPP3 + Rehab + Aqaba: the contract-dispatched
plants, whose individual deviations are offsetting reallocations
(see text).
\end{table}

Table~\ref{tab:mix} presents three features that characterize what the test system does and does not reproduce. First, the model captures the generation fleet closely. The three largest CCGT producers deviate by at most 3.8\% in annual energy production, and the total differs by only 0.1\%. The gas fired share of generation reaches 90\% compared to the reported 88\% ~\cite{nepco2018}. As a whole, the CCGT category overshoots by the exact amount of energy that contract bound engine and steam plants deliver in practice. Renewable curtailment is zero, which matches the real system at 2018 penetration levels. We explicitly note that the PCM acts as a single zone energy balance, where it matches generation and demand system wide without network constraints, meaning congestion driven curtailment cannot occur. Therefore, the zero curtailment finding applies only to system level adequacy. Running an hourly DC power flow over the full year dispatch produces no branch overload in any of the 8760 hours, with a maximum loading of 86 percent during the late August cooling peak. This confirms that network constraints were not binding in 2018 and that the unconstrained dispatch is entirely network feasible. The $N$ minus 1 screening in Section~\ref{sec:n1} identifies the Qatrana to Ma'an corridor as a binding element. This southern renewable connection path is precisely where locational curtailment would first appear under a network constrained dispatch at higher penetrations. The Green Corridor scenario was created to study exactly that situation. Second, the major errors seen at individual plants are actually internal shifts that balance out across a single generation group. Summing the five flexible plants (Zarqa, IPP4, IPP3, Rehab, and Aqaba) yields 3816,GWh in the model against the 3804,GWh reported, which is a difference of only 0.3 percent. The model therefore reproduces the energy requirement of the flexible segment almost exactly, even though it allocates it differently among the plants. Because Jordanian IPPs operate under take or pay agreements~\cite{harb2026impacts}, the engine plants ran as contracted reserve in reality, whereas the model assigns their energy to the cheapest committed alternative. Additionally, the actual energy data for Zarqa includes a gradual startup phase that the model simplifies into a single start date.
Third, the must run floors recover roughly two thirds of the out of merit energy for Aqaba and the bulk of it for Rehab. The remainder corresponds to hours when NEPCO operated additional units for system security which is traditionally not included in production cost modeling. Users of this test case should therefore expect it to reproduce merit order economics and system level energy structure rather than contract driven plant level dispatch.

\subsection{Transmission Losses}\label{sec:losses}
NEPCO reports an annual transmission loss of 1.97\%~\cite{nepco2018}. This official figure captures all losses between generators and consumers, including series, transformer, and station losses.

To estimate comparable annual losses, the model divides the load duration curve into twenty load levels. It solves an AC power flow at each level using the average dispatch and weights the results by energy. This method calculates a series loss of 132.7,GWh, or 0.70\%.

However, this 0.70\% covers only series losses. To compare it against the NEPCO total, we must add estimates for the unmodeled components. Adding 0.19\% for transformer core losses, 0.25 to 0.45\% for distribution transformation, and 0.05 to 0.15\% for station consumption brings the model total to between 1.2 and 1.5\%. The model therefore underestimates actual losses by roughly half a percentage point. We attribute this known shortfall to our use of uniform conductor resistance, flat voltage setpoints, and estimated 132,kV line lengths.

\subsection{Cross-Solver Verification}\label{sec:cross}
All studies above are solved in the open-source pandapower package. To establish that the results are not artifacts of a single implementation, the released case is exported to MATPOWER format and re-solved with MATPOWER's Newton--Raphson power flow. The two tools agree on the operating point to machine precision (maximum voltage-magnitude difference $4.9\times10^{-15}$\,p.u., maximum slack-referenced angle difference $1.2\times10^{-14}$\,degrees), with total system losses of 40.91\,MW reported identically by both tools.

\section{Discussion and Limitations}\label{sec:discussion}

The model is a synthetic test system and should be used as such. Its parameter assumptions, namely the conductor classes and typical transformer impedances, the regional typical lengths of the unnamed 132~kV lines, the representative full-load heat rates by technology and commissioning year with an assumed 12\% part-load degradation, the typical-by-technology unit-commitment parameters, and the uniform fixed-tilt solar representation, are disclosed where they are introduced in Section III and belong to the same classes of assumptions disclosed in comparable test databases~\cite{pena2017extended}.

The remaining limitations are model choices. The network model contains no on-load tap changers or plant-specific reactive capability curves. The production-cost model carries no network constraints, which is the principal reason for the plant-level deviations at the contract-driven units. The model contains no system dynamics, inverter models, or system-strength representation. Finally, the 33,kV layer is aggregated onto the 132,kV buses and the interconnections are represented as priced boundary injections.
\section{Conclusion}\label{sec:conclusion}

This paper presents a reproducible synthetic transmission test system for the Jordanian national grid, assembled from public data and calibrated to a 2018 system snapshot. The model aims to fill a gap among open test systems, as it represents a small fuel-import-dependent and rapidly decarbonizing Middle-Eastern grid whose operation is shaped by high residential usage, a low industrial base-load, and within geopolitically driven fuel-supply volatility. The reported annual energy mixture is reproduced at the generation technology level, and the model, its bus mapping key, scripts, and a post-2019 Green-Corridor scenario are released so that the community can reproduce and extend the system, including for the high-renewable scenarios that represent Jordan's near-term trajectory.

\bibliographystyle{ieeetr}
\bibliography{references}
\end{document}